# Nonlinear transport through a finite Hubbard chain connected to the electrodes

Kamil Walczak [1]

Institute of Physics, Adam Mickiewicz University
Umultowska 85, 61-614 Poznań, Poland

Coherent electronic transport through a molecular device is studied using non-equilibrium Green's function (NEGF) formalism. Such device is made of a short linear wire which is connected to para- and ferromagnetic electrodes. Molecule itself is described with the help of Hückel (tight-binding) model with the electron interactions treated within Hubbard approach, while the coupling to the electrodes is modeled through the use of a broad-band theory. Coulomb interactions within molecular wire are treated by means of the Hartree-Fock (HF) approximation. For the case of asymmetric coupling to paramagnetic electrodes, charging-induced rectification effect in biased molecular devices is discussed as a consequence of Coulomb repulsion. For the system with ferromagnetic electrodes, a significant magnetoresistance (MR) is predicted and its oscillations generated by Coulomb interactions are considered.

## I. Introduction

Quantum transport through small interacting systems has been a subject of current interest. Especially in molecular electronics this problem is still raised, which together with voltage drop across the molecular junction in the presence of applied bias is one of the most important issues [1]. Recent experiments allow to obtain molecular contacts with the help of scanning tunneling microscopy (STM) [2-4] or using the method of mechanically controllable break-junctions (MCB) [5-7]. Molecular junction is composed of two metallic electrodes connected by a molecule (or molecular layer) and operates under the influence of bias voltage. A full knowledge of the conduction mechanism at the molecular scale is not completed yet, but transport properties of considered structures are associated with some quantum effects, such as: quantum tunneling process, quantization of molecular energy levels, discrete nature of electron charge and spin. In general, the current flowing through the molecular junction is influenced by: the internal structure of molecular system, electronic properties of the electrodes near Fermi energy level, character of the coupling between molecule and the electrodes.

Presented studies are focused on the influence of Coulomb interactions on the transport phenomena in nanodevices based on individual molecules. It is shown that the charging effects greatly change current(voltage) I(V) characteristics of molecular junctions. For simplicity and to give a clear physical interpretation of obtained results, we analyze a finite Hubbard chain of size N, which is connected to the electrodes treated within a broad-band theory. In practice, linear carbon-atom chains containing up to $N = 20$ atoms connected at the ends to metal atoms have been synthesized [8] and recognized as ideal one-dimensional wires [9,10]. Proposed model can be used to simulate such system by properly matching of model parameters. The main purpose of this work is to study how the I(V) characteristic depends on the strength of the Coulomb repulsion. Since both electrodes could be made of as well paramagnetic as ferromagnetic materials, transport is considered as spin-dependent. It's a new direction towards molecular spintronics [11-13].

**II. Computational details**

The Hamiltonian of the entire system composed of two electrodes spanned by a molecular wire can be expressed as a three-part sum:

$$H_{tot} = H_{el} + H_{mol} + H_{el-mol}. \tag{1}$$

The first term describes electrons in the electrodes:

$$H_{el} = \sum_{k,\sigma \in \alpha} \varepsilon_{k,\sigma} n_{k,\sigma}, \tag{2}$$

where: $\alpha = L/R$ for the case of left/right (source/drain) electrode, respectively. In the presence of bias voltage, one-electron energies $\varepsilon_{k,\sigma}$ are shifted in the following way: $\varepsilon_{k,\sigma} \to \varepsilon_{k,\sigma} + eV/2$ in the left electrode and $\varepsilon_{k,\sigma} \to \varepsilon_{k,\sigma} - eV/2$ in the right electrode, while chemical potentials of the electrodes are defined through the relations: $\mu_L = \varepsilon_F + eV/2$ and $\mu_R = \varepsilon_F - eV/2$ ($\varepsilon_F$ denotes the equilibrium Fermi level). The second term represents a linear N-atom chain, which is described within the Hubbard model approach:

$$H_{mol} = \sum_{i,\sigma} \varepsilon_{i,\sigma} n_{i,\sigma} - \sum_{i,\sigma} \left( \beta_{i,i+1} c_{i,\sigma}^+ c_{i+1,\sigma} + h.c. \right) + U \sum_i n_{i\uparrow} n_{i\downarrow}, \tag{3}$$

where: $\varepsilon_{i,\sigma}$ is local site energy, $\beta_{i,i+1}$ is the first-neighbor hopping integral, U is the on-site Coulomb interaction between two electrons with opposite spins, while $n_{i,\sigma}$, $c_{i,\sigma}^+$ and $c_{i,\sigma}$ denote the number, creation and annihilation operators for electron on site i with spin $\sigma$. By setting $U = 0$ in Eq.3 one reproduces Hückel (tight-binding) Hamiltonian. The initial condition stems from the assumption that the electric potential between the electrodes varies linearly with the distance (ramp model) [14]. Therefore, the local site energies $\varepsilon_{i,\sigma}$ are shifted due to this voltage ramp: $\varepsilon_{i,\sigma} \to \varepsilon_{i,\sigma} + eV[1 - 2i/(N+1)]/2$. The third term in Eq.1 corresponds to the tunneling process from the electrodes onto the molecule:

$$H_{el-mol} = \sum_{k,\sigma \in \alpha} t_\alpha \left( c_{k,\sigma}^+ c_{i,\sigma} + h.c. \right), \tag{4}$$

where: $t_\alpha$ is hopping integral responsible for the strength of the coupling with the electrode $\alpha$. All the values of energy integrals ($\varepsilon$, $\beta$, U, t) are treated as parameters which can be reasonably modified.

Further analysis are performed within Hartree-Fock (HF) approximation, where the charge occupation number on particular sites are calculated using self-consistent procedure. The HF problem is associated with a simplification of molecular Hamiltonian (2), which can be rewritten in the form:

$$H_{mol}^{HF} = \sum_{i,\sigma} \widetilde{\varepsilon}_{i,\sigma} n_{i,\sigma} - \sum_{i,\sigma} \left( \beta_{i,i+1} c_{i,\sigma}^+ c_{i+1,\sigma} + h.c. \right), \tag{5}$$

with the local site energy given by:

$$\widetilde{\varepsilon}_{i,\sigma} = \varepsilon_{i,\sigma} + U \langle n_{i,\overline{\sigma}} \rangle, \tag{6}$$

Occupation number of the electrons on each site for particular voltages (nonequilibrium case) is determined self-consistently using the Keldysh formalism [15]:

$$\langle n_{i,\sigma} \rangle = -\frac{i}{2\pi} \int_{-\infty}^{+\infty} d\omega G^{<}_{i\sigma,i\sigma}(\omega), \qquad (7)$$

The lesser Green function $G^{<}$ can be obtained from the Dyson equation and expressed in the general form as:

$$G^{<}_{i\sigma,j\sigma} = \sum_{i',j'} G^{r}_{i\sigma,i'\sigma} \Sigma^{<}_{i'\sigma,j'\sigma} G^{a}_{j'\sigma,j\sigma}. \qquad (8)$$

The superscripts r and a denote the retarded and advanced Green functions, respectively:

$$G^{r}(\omega) = \left[ J\omega - H^{HF}_{mol} - \Sigma^{r} \right]^{-1} \qquad (9)$$

and $G^{a} = [G^{r}]^{*}$ (here J denotes the unit matrix of the dimension equal to molecular Hamiltonian $N \times N$). The lesser self-energy can be written as follows:

$$\Sigma^{<}_{i\sigma,j\sigma}(\omega) = 2i\delta_{i,j} \left[ \delta_{i,1} \Delta_{L\sigma}(\omega) f_{L}(\omega) + \delta_{i,N} \Delta_{R\sigma}(\omega) f_{R}(\omega) \right], \qquad (10)$$

where $f_{\alpha}$ is Fermi distribution function in the $\alpha$ electrode. In our case:

$$\Delta_{\alpha\sigma}(\omega) = \pi t_{\alpha}^{2} \rho_{\alpha\sigma}, \qquad (11)$$

where $\rho_{\alpha\sigma}$ is local density of states (at the Fermi energy level) for the electrons with spin $\sigma$ in the $\alpha$ electrode. The retarded and advanced self-energy functions are given by:

$$\Sigma^{r}_{i\sigma,j\sigma}(\omega) = \delta_{i,j} \left[ \delta_{i,1} \left[ \Lambda_{L\sigma}(\omega) - i\Delta_{L\sigma}(\omega) \right] + \delta_{i,N} \left[ \Lambda_{R\sigma}(\omega) - i\Delta_{R\sigma}(\omega) \right] \right] \qquad (12)$$

and $\Sigma^{a} = [\Sigma^{r}]^{*}$. The real and imaginary terms of the self-energy components are closely connected to each other through the Hilbert transform [16]:

$$\Lambda_{\alpha\sigma}(\omega) = \frac{1}{\pi} P \int_{-\infty}^{+\infty} dz \frac{\Delta_{\alpha\sigma}(z)}{\omega - z}, \qquad (13)$$

where P is the Cauchy principal value. For the sake of simplicity, in further analysis we make an assumption that local density of states in both electrodes is constant over an energy bandwidth and zero otherwise. This case of dispersionless coupling to the electrodes is commonly used in the literature and is usually sufficient in describing broad-band metals.

The current flowing through the device can be computed from the time evolution of the occupation number for electrons in the left (or equivalently right) electrode and can be expressed by the lesser Green function [15]:

$$I(V) = -e\frac{d}{dt}\langle N_{\alpha} \rangle = \frac{e}{h} \sum_{i,k,\sigma} t_{\alpha} \int_{-\infty}^{+\infty} d\omega \left[ G^{<}_{i\sigma,k\sigma}(\omega) + c.c. \right], \qquad N_{\alpha} = \sum_{k,\sigma \in \alpha} n_{k,\sigma}. \qquad (14)$$

After applying the Dyson equation, current formula can be written with the help of the retarded Green function:

$$I(V) = \frac{4e}{h} \sum_\sigma \int_{-\infty}^{+\infty} d\omega [f_L(\omega) - f_R(\omega)] \Delta_{L\sigma}(\omega) \Delta_{R\sigma}(\omega) |G^r_{1\sigma,N\sigma}(\omega)|^2. \qquad (15)$$

In our calculations, we have assumed that transport process is purely coherent and elastic. It means that the current conservation rule is fulfilled on each site and for any energy $\omega$.

**III. Non-magnetic case**

As an example we take into consideration a linear carbon-atom chain containing $N = 4$ atoms sandwiched between two paramagnetic electrodes [14]. Molecular description itself includes only $\pi$-electrons of hydrocarbons (and is based on the assumption of one $2p_z$-basis function for each carbon atom), while the coupling to the electrodes is treated within a broad-band theory. This is a test case simple enough to analyze all the essential physics in detail and to compare obtained results with results already known in the literature [14]. Therefore, we take the following energy parameters (given in eV): $\varepsilon = 0$ (the reference energy), $\beta = 2.4$. Fermi energy for organic junctions is usually fixed in between the highest occupied molecular orbital (HOMO) and the lowest unoccupied molecular orbital (LUMO). In this work we assume that Fermi level is fixed 1 eV below the LUMO level of isolated molecule ($\varepsilon_F = 0.483$). The choice of the particular value of Coulomb integral is somewhat arbitrary, while temperature of the system is set at $T = 293$ K (however, obtained results are not particularly sensitive to temperature). Both effects of the temperature and the molecule-to-electrodes coupling are associated with smoothing all the transport characteristics.

Molecular chain symmetrically connected to the electrodes ($\Delta_L = \Delta_R = 0.025$) represents the experimental situation in which molecules are inserted into break junction [5-7]. The numerical results of the I(V) curves for this case are shown in Fig.1. Computations for $U = 0$ (Hückel approach with the potential ramp) represent two distinct current jumps separated by the current plateau (in the voltage range $0 < V < 6$). It is clear that in our model, electron and hole transport processes are analyzed separately, since the Fermi level of the system is fixed closer to the LUMO level (not in the middle of the HOMO-LUMO gap). The first step in the I(V) curve is due to the electron conduction through the LUMO level, while the second one is associated with the hole conduction through the HOMO level. The two steps have the same intensity, because of the electron-hole symmetry of the model.

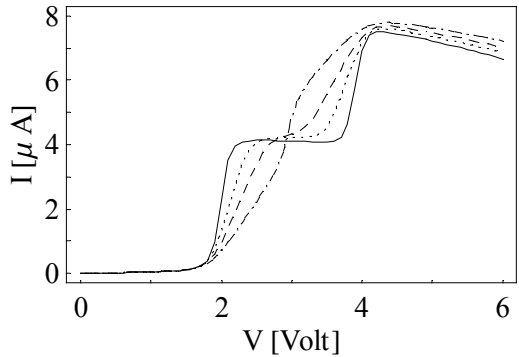

Fig.1 The evolution of the $I(V) = -I(-V)$ curves for various Coulomb integrals: $U = 0$ (solid line), $U = 1$ (dotted line), $U = 2$ (dashed line) and $U = 4$ (dashed-dotted line). The other parameters of the model: $N = 4$, $\varepsilon = 0$, $\beta = 2.4$, $\Delta_L = \Delta_R = 0.025$, $\varepsilon_F = 0.483$.

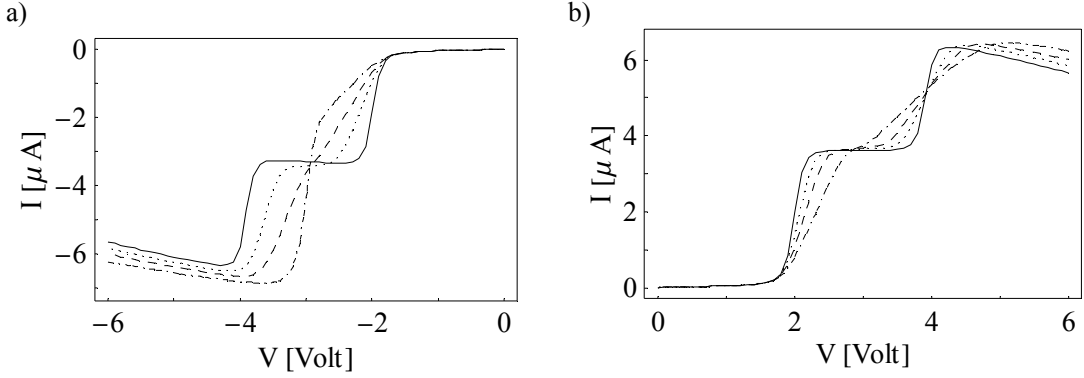

Fig.2 The evolution of the $I(V)$ curves for various Coulomb integrals: $U = 0$ (solid line), $U = 1$ (dotted line), $U = 2$ (dashed line) and $U = 4$ (dashed-dotted line). The other parameters of the model: $N = 4$, $\varepsilon = 0$, $\beta = 2.4$, $\Delta_L = 0.035$, $\Delta_R = 0.015$, $\varepsilon_F = 0.483$. (a) Results for voltage range $-6 < V < 0$ and (b) for voltage range $0 < V < 6$.

Coulomb repulsion leads to pinning of the molecular self-consistent levels to the chemical potentials of the source and drain electrodes, and resonant tunneling steps in the I-V curves are smoothed due to the mentioned effect [17,18]. That's why typical sequence of irregularly placed steps in the I(V) dependence (related to electronic structure of the molecular wire for $U = 0$) is replaced by fairly smooth I(V) function with the current steps of a finite slope (transformed into wide peaks in the corresponding differential conductance for $0 < U < \beta$). In general, the slope of the current step decreases with the increase of the U-parameter. The level pinning extends over finite voltage regions of the range of $U/e$ and this continuous shift of the levels is caused by the Hartree contributions $U\langle n_{i\bar{\sigma}}\rangle$ to the local site energies. Therefore, in the limit of strong Coulomb interactions ($U \geq \beta$) two current steps join together into one step at $V \approx 3$ (approximately doubled in height). It means that in this regime, electron and hole conduction partly merges (in transport process participates simultaneously both the HOMO and the LUMO levels). However, it is well-known that for the case of $U \geq 2\beta$, the iterative self-consistent procedure diverges [19]. Moreover, the I(V) dependence exhibits a negative differential resistance (NDR) behavior, seen for higher voltages ($V > 4$) independently from the strength of the Coulomb integral U. The NDR effect is a consequence of a voltage drop inside the molecule [20].

Molecular chain asymmetrically connected to the electrodes ($\Delta_L = 0.035$, $\Delta_R = 0.015$) represents the experimental situation in which scanning tunneling microscope (STM tip) is used in order to receive molecular contact [2-4]. The numerical results of the I(V) curves for this case are presented in Fig.2, where those characteristics reflect all the features discussed earlier. A new qualitative result is associated with the rectification effect, i.e. phenomenon in which the magnitude of the junction current depends on the polarity of bias voltage. In order to describe such behavior, the so-called rectification ratio (RR) for particular bias voltage is defined with the help of the following relation:

$$RR(V) = |I(V)/I(-V)|. \qquad (16)$$

It can be observed in Fig.3 that I(V) dependences for molecule asymmetrically coupled to the electrodes show rectification behavior always after the first current step. Furthermore, such effect is induced by Coulomb interactions within the molecule (for $U > 0$), since diode-like behavior is absent in the case of $U = 0$.

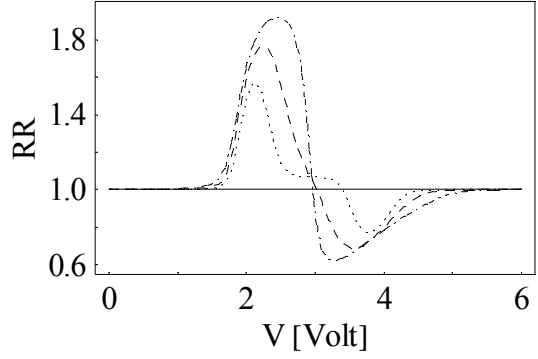

Fig.3 The voltage dependence of the rectification ratio for various Coulomb integrals: $U = 0$ (solid line), $U = 1$ (dotted line), $U = 2$ (dashed line) and $U = 4$ (dashed-dotted line). The other parameters of the model: $N = 4$, $\varepsilon = 0$, $\beta = 2.4$, $\Delta_L = 0.035$, $\Delta_R = 0.015$, $\varepsilon_F = 0.483$.

General tendency is following: the higher values of U-parameter, the stronger rectification effect is generated. It can be also concluded that in the realm of electron conduction $RR > 1$ (and therefore $|I(V)| > |I(-V)|$), while in the realm of hole conduction $RR < 1$ (and therefore $|I(V)| < |I(-V)|$).

## IV. Spin-dependent transport

In the presence of magnetic impurities or upon application of ferromagnetic electrodes, spin-dependent transport through molecular junctions is expected [11-13]. Phenomena of that type are of great interest due to potential applications in near-future nanoelectronics. In this section we analyze a linear carbon-atom chain containing $N = 4$ atoms connected at both ends to ferromagnetic electrodes, where ferromagnetism is taken into account simply through different densities of states ($\rho_{\alpha\uparrow}$ and $\rho_{\alpha\downarrow}$) for the electrons with spin $\sigma$ in the $\alpha$ electrode. Let us assume for that purpose: $\Delta_{\alpha\uparrow} = 0.035$ and $\Delta_{\alpha\downarrow} = 0.015$ for $\alpha = L, R$.

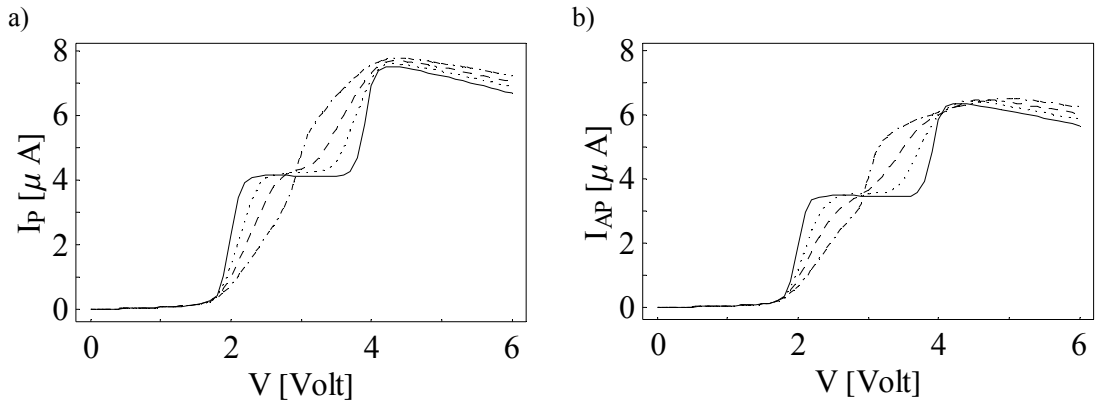

Fig.4 The evolution of the $I(V) = -I(-V)$ curves for various Coulomb integrals: $U = 0$ (solid line), $U = 1$ (dotted line), $U = 2$ (dashed line) and $U = 4$ (dashed-dotted line). Parameters of the model: $N = 4$, $\varepsilon = 0$, $\beta = 2.4$, $\Delta_{L\uparrow} = \Delta_{R\uparrow} = 0.035$, $\Delta_{L\downarrow} = \Delta_{R\downarrow} = 0.015$, $\varepsilon_F = 0.483$.
(a) Results for parallel and (b) antiparallel alignment of the spin polarization in the electrodes.

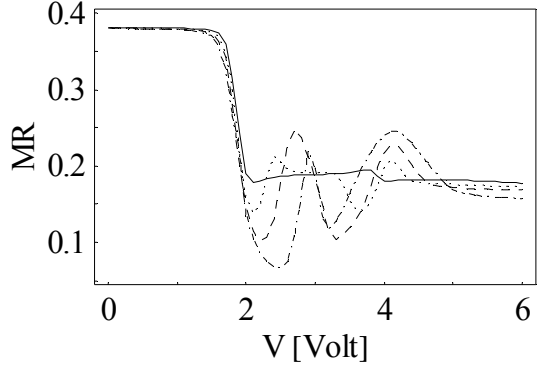

Fig.5 The voltage dependence of the magnetoresistance $MR(V) = MR(-V)$ for various Coulomb integrals: $U = 0$ (solid line), $U = 1$ (dotted line), $U = 2$ (dashed line) and $U = 4$ (dashed-dotted line). The other parameters of the model: $N = 4$, $\varepsilon = 0$, $\beta = 2.4$, $\Delta_{L\uparrow} = \Delta_{R\uparrow} = 0.035$, $\Delta_{L\downarrow} = \Delta_{R\downarrow} = 0.015$, $\varepsilon_F = 0.483$.

By applying an external magnetic field one can change the orientation of magnetizations in the electrodes. Moreover, the magnetoresitance (MR) can be defined as a relative difference of the current in the parallel (P) and antiparallel (AP) configuration of the spin polarization alignment in the electrodes [13]:

$$MR(V) = [I_P(V) - I_{AP}(V)]/I_{AP}(V). \qquad (17)$$

Numerical results for I(V) calculations in the case of P and AP alignments are shown in Fig.4. Here I(V) functions once more reveal all the features discussed in the previous section, although current for P configuration reaches higher values than current for AP configuration ($I_P > I_{AP}$). From Fig.5 we can see that until the first current step ($V \approx 2$) MR is stabilized on the same level $MR \approx 0.38$ independently from the strength of Coulomb interactions. However, for higher voltages ($V > 2$) MR coefficient is smaller and oscillates around the value $MR \approx 0.19$. General tendency is following: the higher values of U-parameter, the stronger oscillations of magnetoresistance are expected.

## V. Concluding remarks

Summarizing, nonequilibrium spin-dependent transport calculations were performed in the molecular device made of molecular wire attached to para- and ferromagnetic electrodes. Molecular system was treated as a linear Hubbard chain (at the Hartree-Fock level), while the coupling to the electrodes was described within a broad-band theory. It was shown that charge transfer through molecules strongly depends on the Coulomb repulsion, leading to continuous I-V dependences in the weak-coupling limit (like in the most experimental data). For the case of asymmetric coupling to paramagnetic electrodes, charging-induced rectification effect can be observed. Until now, a number of molecular diodes have been proposed and all of those systems can be classified into four categories: donor/σ-bridge/acceptor molecules [21], donor/π-bridge/acceptor zwitterionic molecules [22], nonalternant hydrocarbons [23] and π-conjugated molecules asymmetrically coupled to two electrodes [24]. In this article we have shown that asymmetric coupling itself is maybe not enough to generate rectification effect, but together with Coulomb interactions (within molecular wire) can lead to rectifying behavior. Such conclusion is in good agreement with previous result obtained in slightly different approach to the problem of electronic transport in molecular devices (i.e. generalized Breit-Wigner formula) [25]. For the system with ferromagnetic electrodes, the current for the P configuration is higher than in the AP one, leading to a positive magnetoresistance in the system. Moreover, oscillations of the magnetoresistance are seen after the first current step onto the I-V characteristics, which are generated by Coulomb interactions. This result belongs to the most intriguing in our work.

However, it should be also stressed that the method presented in this work is based on few drastic assumptions and therefore all the obtained results should be considered as qualitative only. For instance, in HF approach all the many-body effects are omitted and electronic correlations are neglected. Besides, the HF approximation is well-known to overestimate the role of electron repulsion. Since in our model we do not include any inter-site interaction parameter, we believe that both effects can somehow compensate. Anyway, studies on the influence of long-range Coulomb interactions on transport characteristics are in progress.

In the low-temperature limit (below the so-called Kondo temperature $T < T_K$) one can also expect some other effects associated with charge and spin fluctuations, since molecule is successively charge and discharged during the electronic transport through the junction. For odd number of electrons in the molecular dot, it leads to the Kondo resonance and formation of the peak in the electron density of states. At equilibrium the Kondo peak is aligned with the Fermi level of the electrodes. The peak height increases logarithmically with reduction of the temperature [26], resulting in perfect transmission at $T = 0$ K (where conductance is equal to $2e^2/h \approx 77.5$ μS) [27]. Moreover, in the presence of bias voltage the Kondo level is split into two resonances which are pinned to chemical potentials of the source and drain electrodes [28]. The split Kondo peaks are suppressed by a finite lifetime due to dissipative transitions in which electrons are transferred from one electrode to another. Upon application of an external magnetic field $B$, the Kondo peaks are additionally shifted from the chemical potentials by the ordinary Zeeman splitting $g\mu_B B$ (where $g \approx 2$ is the $g$-factor of the molecule, while $\mu_B \approx 57.5$ μeV is the Bohr magneton), but in opposite directions for each spin [26,29].

Concluding, the strong interest of molecular devices stems from their: small sizes, quantum nature of electrical conduction, theoretically inexhaustible possibilities of structural modifications of the molecules, relatively low costs and easiness in obtaining layer-based molecular junctions (due to self-assembly features of organic molecules). Among the most important tasks in molecular electronics we can enumerate: fabrication of molecular junctions, understanding of the mechanisms of the current flowing through such devices, determination of the main factors that control transport phenomena in molecular systems, and eventually connection of individual devices into a properly working integrated circuit. The final goal of molecular electronics might be construction of supercomputer with molecular processor that could have extraordinary parameters (frequency of THz and memory of GB) [30].

**Acknowledgments**

The author is grateful to T. Kostyrko and B. Bułka for many valuable discussions.

# References

¹ E-mail: walczak@amu.edu.pl
[1] A. Nitzan, and M. A. Ratner, Science **300**, 1384 (2003).
[2] A. Stabel, P. Herwig, K. Müllen, and J. P. Rabe,
    Angew. Chem. Int. Ed. Engl. **34**, 1609 (1995).
[3] W. Tian, S. Datta, S. Hong, R. Reifenberger, J. I. Henderson, and C. P. Kubiak,
    J. Chem. Phys. **109**, 2874 (1998).
[4] B. Xu, and N. J. Tao, Science **301**, 1221 (2003).
[5] M. A. Reed, C. Zhou, C. J. Muller, T. P. Burgin, and J. M. Tour,
    Science **278**, 252 (1997).
[6] C. Kergueris, J.-P. Bourgoin, S. Palacin, D. Esteve, C. Urbina, M. Magoga,
    and C. Joachim, Phys. Rev. B **59**, 12505 (1999).
[7] J. Reichert, R. Ochs, D. Beckmann, H. B. Weber, M. Mayor, and H. v. Löhneysen,
    Phys. Rev. Lett. **88**, 176804 (2002).
[8] G. Roth, and H. Fischer, Organometallics **15**, 5766 (1996).
[9] N. D. Lang, and Ph. Avouris, Phys. Rev. Lett. **81**, 3515 (1998).
[10] N. D. Lang, and Ph. Avouris, Phys. Rev. Lett. **84**, 358 (2000).
[11] E. G. Emberly, and G. Kirczenow, Chem. Phys. **281**, 311 (2002).
[12] W. I. Babiaczyk, and B. R. Bułka, Phys. Stat. Sol. (a) **196**, 169 (2003).
[13] W. I. Babiaczyk, and B. R. Bułka, J. Phys.: Condens. Matter **16**, 4001 (2004).
[14] V. Mujica, M. Kemp, A. Roitberg, and M. Ratner, J. Chem. Phys. **104**, 7296 (1996).
[15] H. Haug, and A.-P. Jauho, *Quantum Kinetics in Transport and Optics
     of Semiconductors*, Springer-Verlag, Berlin Heildelberg 1998.
[16] V. Mujica, M. Kemp, and M. A. Ratner, J. Chem. Phys. **101**, 6849 (1994);
     *ibid*. **101**, 6856 (1994).
[17] M. Paulsson, and S. Stafström, Phys. Rev. B **64**, 035416 (2001).
[18] T. Kostyrko, and B. R. Bułka, Phys. Rev. B **67**, 205331 (2003).
[19] G. Treboux, J. Phys. Chem. B **104**, 9823 (2000).
[20] Y. Xue, S. Datta, S. Hong, R. Reifenberger, J. I. Henderson, and C. P. Kubiak,
     Phys. Rev. B **59**, R7852 (1999).
[21] A. Aviram, and M. A. Ratner, Chem. Phys. Lett. **29**, 277 (1974).
[22] S. Martin, J. R. Sambles, and G. J. Ashwell, Phys. Rev. Lett. **70**, 218 (1993).
[23] G. Treboux, P. Lapstun, and K. Silverbrook, J. Phys. Chem. B **102**, 8978 (1998).
[24] C. Zhou, M. R. Deshpande, M. A. Reed, L. Jones II, and J. M. Tour,
     Appl. Phys. Lett. **71**, 611 (1997).
[25] K. Walczak, Physica E (2004) in press. DOI: 10.1016/j.physe.2004.08.102.
[26] J. Park *et al*., Nature **417**, 722 (2002).
[27] T. K. Ng, and P. A. Lee, Phys. Rev. Lett. **61**, 1768 (1988).
[28] Y. Meir, N. S. Wingreen, and P. A. Lee, Phys. Rev. Lett. **70**, 2601 (1993).
[29] W. J. Liang *et al*., Nature **417**, 725 (2002).
[30] Y. Wada, Pure Appl. Chem. **71**, 2055 (1999).